\documentclass[aos,preprint]{imsart}

\RequirePackage[OT1]{fontenc}
\RequirePackage{amsthm,amsmath,amssymb}
\RequirePackage[numbers]{natbib}

\usepackage{mathrsfs}
\usepackage{graphicx}
\usepackage{verbatim}
\usepackage{float}

\newtheorem{defin}{Definition}

\newtheorem{lem}{Lemma}
\newtheorem{thm}{Theorem}

\newcommand{\diag}{\mathrm{diag}}

\newcommand{\beq}{\begin{equation}}
\newcommand{\eeq}{\end{equation}}

\begin{document}

\begin{frontmatter}
\title{A Geometrical Approach to  Topic Model Estimation}
\runtitle{Geometrical Approach for Topic Models}

\begin{aug}
\author{\fnms{Zheng Tracy} \snm{Ke}\ead[label=e1]{zke@galton.uchicago.edu}}

\runauthor{Z. Ke}

\affiliation{University of Chicago}

\address{Z. Ke\\
Department of Statistics\\
University of Chicago\\
\printead{e1}\\
\phantom{E-mail: zke@galton.uchicago.edu}}

\end{aug}

\begin{abstract}
In the probabilistic topic models, the quantity of interest---a low-rank matrix consisting of topic vectors---is hidden in the text corpus matrix,  masked by noise,  and the Singular Value Decomposition (SVD) is a potentially useful tool for learning such a low-rank matrix. However, the connection between this low-rank matrix and the singular vectors of the text corpus matrix  are usually complicated and hard to spell out, so how to use SVD for learning topic models faces challenges.  In this paper, we overcome the challenge by revealing a surprising 
insight:  there is a low-dimensional {\it simplex} structure  which can be viewed as a bridge between the low-rank matrix of interest and   
the SVD of the text corpus matrix, and allows us 
to conveniently reconstruct the  former using the latter. 
Such an insight motivates a new SVD approach to learning topic models, which we analyze with  delicate random matrix theory and derive the rate of convergence.  We  support our methods and theory numerically, using both  simulated data and real data.  
\end{abstract}

\end{frontmatter}

\section{Introduction}

The topic models are used in a wide variety of  application areas including but not limited to digital humanities,  computational social science,  e-commerce, and government science policy \cite{blei2012probabilistic}.  The {\it probabilistic Latent Semantic Indexing (pLSI)} is a popular version 
of the topic model  \cite{Hofmann1999}. For a data set with $n$ documents on a vocabulary of $p$ words, we let $D \in \mathbb{R}^{p,n}$ be the so-called {\it text corpus matrix}, where column $j$ of $D$ is the observed fractions of all $p$ words in document $j$,  $1 \leq j \leq n$.      
Write  
\[ 
D = D_0 + Z, \qquad  \mbox{where}\quad D_0 \equiv E[D] \quad \mbox{and}\quad  Z \equiv D - D_0.  
\]   
Despite that there are a large number of words and a large number of 
documents, there are relatively few {\it topics}. Letting $K$ be the number of topics, the pLSI assumes 
each topic $k$ is a distribution over $p$ words, represented by the probability mass function $A_k=[A_k(1),\cdots,A_k(p)]^T$, and each document $j$ is represented as a list of mixing proportions $W_j=[W_j(1),\cdots,W_j(K)]^T$ over $K$ topics. For document $j$, each word in the text is  
independently drawn from the vocabulary as follows: it first draws a topic so that with probability $W_j(k)$ the underlying topic is $k$, $1\leq k\leq K$; given that the underlying topic is $k$, it draws a word from the vocabulary according to the distribution $A_k$. As a result, 
\beq \label{Mod1}
D_0 = A W,  \qquad A  = [A_1, \ldots, A_K]  \in \mathbb{R}^{p, K}, \;\; W = [W_1, \ldots, W_n]  \in \mathbb{R}^{K,n}. 
\eeq
Frequently,  from a practical perspective,    $\mathrm{rank}(AW) \leq K \ll  \min\{n, p\}$.

%
%
%
%

The primary interest of the paper is to use the $p \times n$ text corpus matrix $D$ to estimate the $p \times K$ topic matrix $A$.   
To this end,  a well-known approach to topic model learning is the so-called Latent Dirichlet Allocation (LDA) method  \cite{Blei}.  The approach is practically popular,  but it is hard to analyze theoretically, and it remains unclear what conditions are required for such an approach 
to provide  the desired theoretical guarantee. 
Another popular approach is the the so-called Nonnegative Matrix Factorization (NMF) approach. This approach is theoretically more tractable \cite{arora2013practical, Ge}, but it does not explicitly explore the low-rank feature of $D_0$ in the pLSI model,  as one might have hoped.

%
%
%
%

Our proposal is to use a {\it Singular Value Decomposition (SVD)} approach, as a direct response to the low-rank structure of $D_0$. The main challenge we face here is that, the connection between the quantity of interest---the topic matrix $A$---and the SVD is indirect and opaque.  
This partially explains why the SVD-based methods are not as successful as one might have expected:  literature works either choose to shun from the target (e.g., \cite{azar2001spectral, kleinberg2003convergent, kleinberg2008using} choose to estimate the column span of $A$, instead of the matrix $A$ itself) or choose to impose additional assumptions (e.g., \cite{bansal2014provable, Papadimitriou2000217}).  
Alternatively, one may first use SVD to get a low-rank approximation of the text corpus matrix $D$ and then apply LDA-based or NMF-based methods to the low-rank proxy. 
Unfortunately, this does not work very well either, because the low-rank proxy (which is not even a nonnegative matrix) significantly violates 
the conditions required for the success of either the LDA-based methods or the NMF-based methods. 

%
%
%
%
%
%
%

Our strategy is to get to the bottom of the SVD approach, and to make the  connection between the topic matrix $A$ (quantity of interest) and the SVD (which used to be opaque)  transparent. 

The main surprise of the paper is as follows. Let $\hat{R} \in  \mathbb{R}^{p, K-1}$ be the {\it matrix of entry-wise ratios} between the first $K$ left singular vectors of $D$ (formed by dividing the $(k+1)$-th  left singular vector of $D$ by the first left singular vector of $D$, coordinate-wise, $1 \leq k \leq K-1$; see details below). We discover that there is a simplex which lives in dimension $(K-1)$ and has $K$ vertices, such that, if we view each row of $\hat{R}$ as a point in $\mathbb{R}^{K-1}$, then in the idealized case where 
there is no noise (i.e., $Z = 0$), we have the following observation.   
\begin{itemize} 
\item If word $i$ is an {\it anchor} word, row $i$ of $\hat{R}$ falls exactly on one of the vertices of the 
simplex; throughout this paper, we follow the convention in the literature to call word $i$ an anchor word if $A_k(i)\neq 0$ for exactly one $k$.  
\item Otherwise, row $i$ of $\hat{R}$ falls into the interior (or interior of an edge or a face) of the simplex; moreover, row $i$ is a weighted linear combination of all $K$ vertices, where the weights 
are intrinsically linked to row $i$ of $A$, and knowing the weights means knowing row $i$ of $A$. 
\end{itemize} 
Such a simplex structure, together with $\hat{R}$,  can then be used to reconstruct all $K$ vertices of the simplex and all the weights aforementioned; once we know all the vertices and weights, the matrix $A$ can be conveniently recovered.  The connection between $A$ and the SVD is now transparent.

While the above is for the idealized case, the idea continues to work in the real case where the noise presents,  provided that some regularity conditions hold. Of course, to get all these ideas work, we need non-trivial efforts, and especially several innovations in methods and in theory; see details below.

It is worthy to mention that our simplex is different from those in the literature (e.g., the simplicial cone in \cite{donoho2003does}), which is not based on the SVD. 

The remaining part of the paper is organized as follows. In Section~\ref{sec:geometry}, we describe the simplex in the idealized case, and illustrate how to use it to reconstruct the topic matrix $A$. In Section~\ref{sec:method}, we extend the approach to the real case and propose a new method for topic estimation; the main challenge is how to estimate vertices of a simplex from noisy data, and we address it by a vertices estimation algorithm. In Section~\ref{sec:theory}, we derive the $\ell_1$ estimation error of our method in an asymptotic framework. Sections~\ref{sec:numeric} and \ref{sec:discuss} contain numerical results and discussions, respectively, and Section~\ref{sec:proof} contains the proof of the key lemma, Lemma~\ref{lem:key}. Other proofs are relegated to the supplementary material.

\section{The simplex structure (idealized case)}  \label{sec:geometry}
Recall that $D = D_0 + Z$, where $D_0 = AW$.   
Let $\hat{\sigma}_1 > \hat{\sigma}_2 > \ldots > \hat{\sigma}_K$ and  
$\sigma_1 >  \sigma_2 > \ldots >  \sigma_K$ be the first $K$ singular values of $D$ and $D_0$ respectively, and let  $\hat{\xi}_1, \hat{\xi}_2, \ldots, \hat{\xi}_K$ and $\xi_1, \xi_2, \ldots, \xi_K$ be the corresponding left singular vectors.  Write $\hat{\Xi} = [\hat{\xi}_1, \hat{\xi}_2, \ldots, \hat{\xi}_K]$ and $\Xi = [\xi_1, \xi_2, \ldots, \xi_K]$. 

In this section, we consider an idealized case where $\Xi$ is given, and look for an approach (a) that  can be used to conveniently reconstruct $A$ using $\Xi$ (in the idealized case where $\Xi$ is given), and (b) that can be easily extended to the {\it real} case where $\hat{\Xi}$ is available (but $\Xi$ is not available, of course). 


\begin{defin}
For any vector $d \in \mathbb{R}^n$, $\diag(d)$ denotes the $n \times n$ diagonal matrix where the $i$-th diagonal entry is the $i$-th entry of $d$, $1 \leq i \leq n$. 
\end{defin}

By linear algebra, there exists a $K \times K$ matrix $V$ such that 
\beq \label{defineV}
\Xi = A V.
\eeq
Let $a_i\in\mathbb{R}^K$ be the $i$-th row of $A$, $1\leq i\leq p$. 
Let $h \in \mathbb{R}^p$ be the vector such that $h(i)   = \|a_i\|_1$,  and let $\tilde{a}_i = a_i /\|a_i\|_1$, $1 \leq i  \leq p$. Let $\tilde{A} = [\tilde{a}_1, \tilde{a}_2, \ldots, \tilde{a}_p]^T$. By these definitions,
\beq 
\Xi = \diag(h)\cdot \tilde{A}V, \qquad h= (\|a_1\|_1,\cdots,\|a_p\|_1)^T,\quad \tilde{A}= [\diag(h)]^{-1}A. 
\eeq
Let $v_1,\cdots, v_K\in\mathbb{R}^K$ be the rows of $V$. Define 
${\cal S}_K$ as the simplex in $\mathbb{R}^{K}$ with $v_1, v_2, \ldots, v_K$ being the vertices. Since each $\tilde{a}_i$ is a weight vector,\footnote{A weight vector is such that all its entries are nonnegative and sum up to $1$.} viewing each row of $\tilde{A} V$ as a point in $\mathbb{R}^{K}$,  the following lemma is a direct observation.   

\begin{lem}\label{lem1}
If word $i$ is an anchor word, then row $i$ of $\tilde{A}V$ coincides with one of the $K$ vertices of ${\cal S}_k$. If word $i$ is not an anchor word, then row $i$ of $\tilde{A} V$ falls into the interior of ${\cal S}_K$ (or the interior of an edge/face of ${\cal S}_K$); moreover, 
it is also a convex combination of $v_1, v_2, \ldots, v_K$, with $\tilde{a}_i$ being the weight vector,  $1 \leq i \leq p$.   
\end{lem}

\begin{figure}[t]
\centering
\includegraphics[width = .4\textwidth, height=.275\textwidth, trim=10 0 0 10, clip=true]{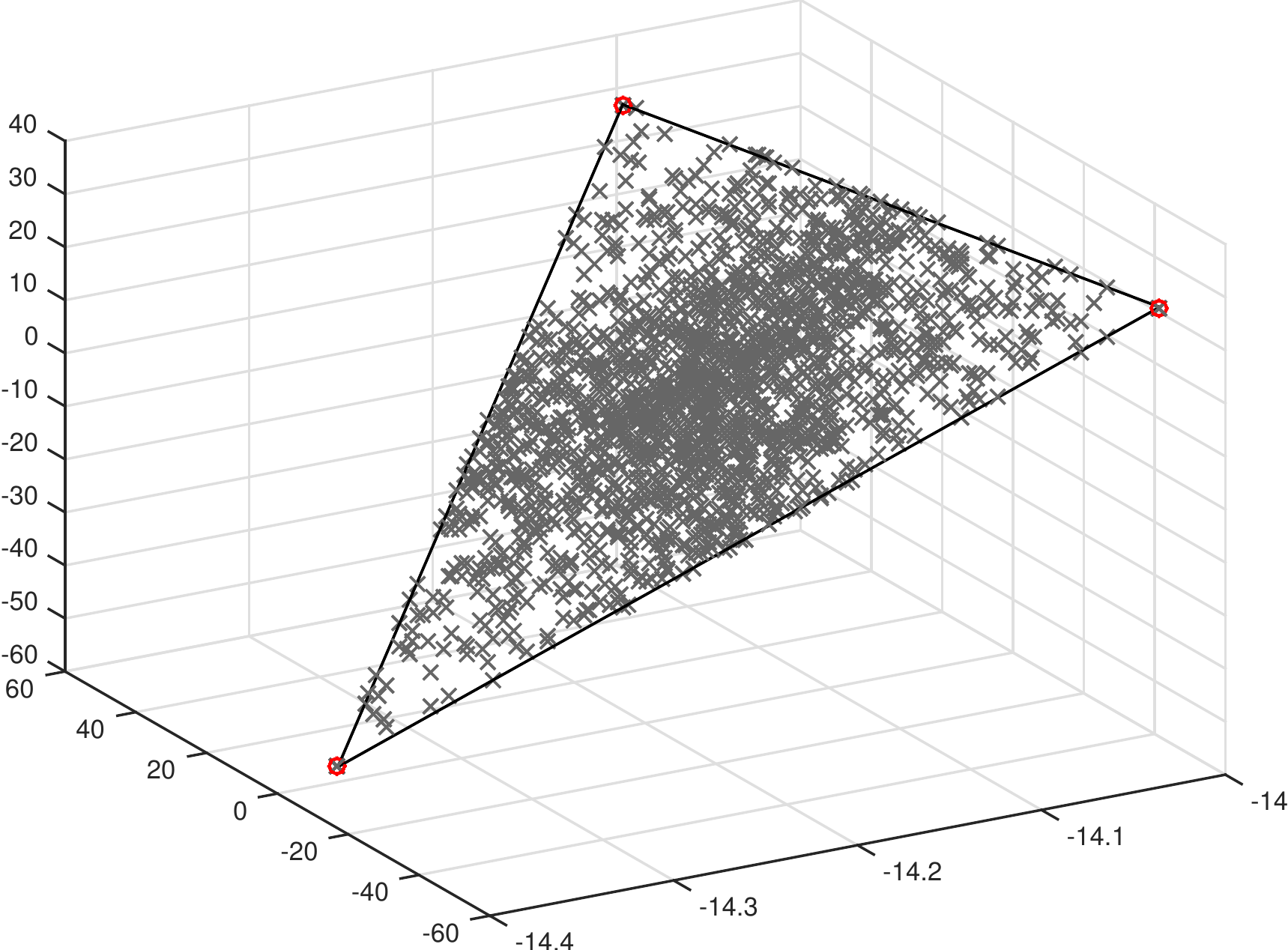}
\includegraphics[width = .4\textwidth, height=.28\textwidth, trim=30 40 10 0, clip=true]{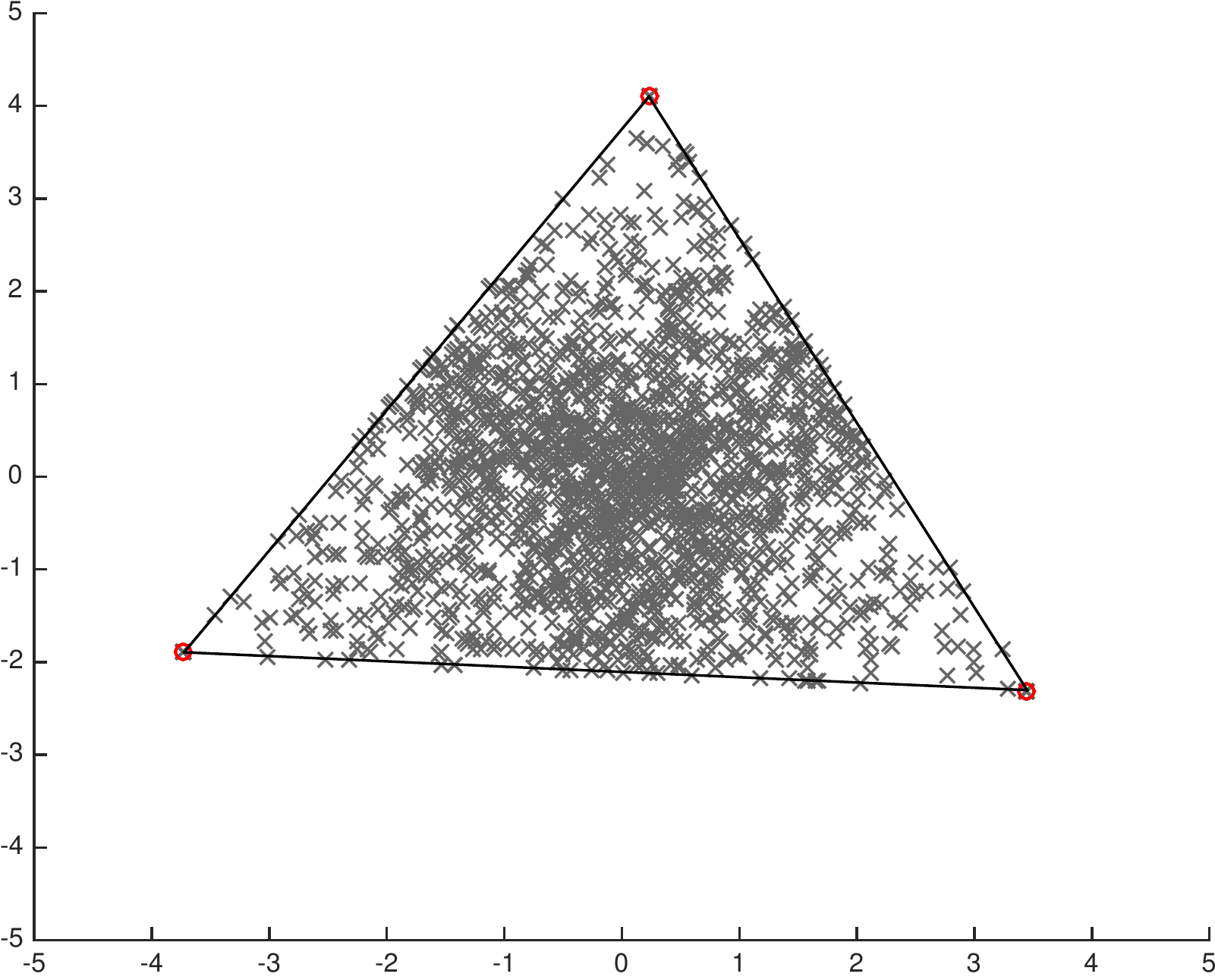}\\
\includegraphics[width = .4\textwidth, height=.28\textwidth, trim=30 40 10 0, clip=true]{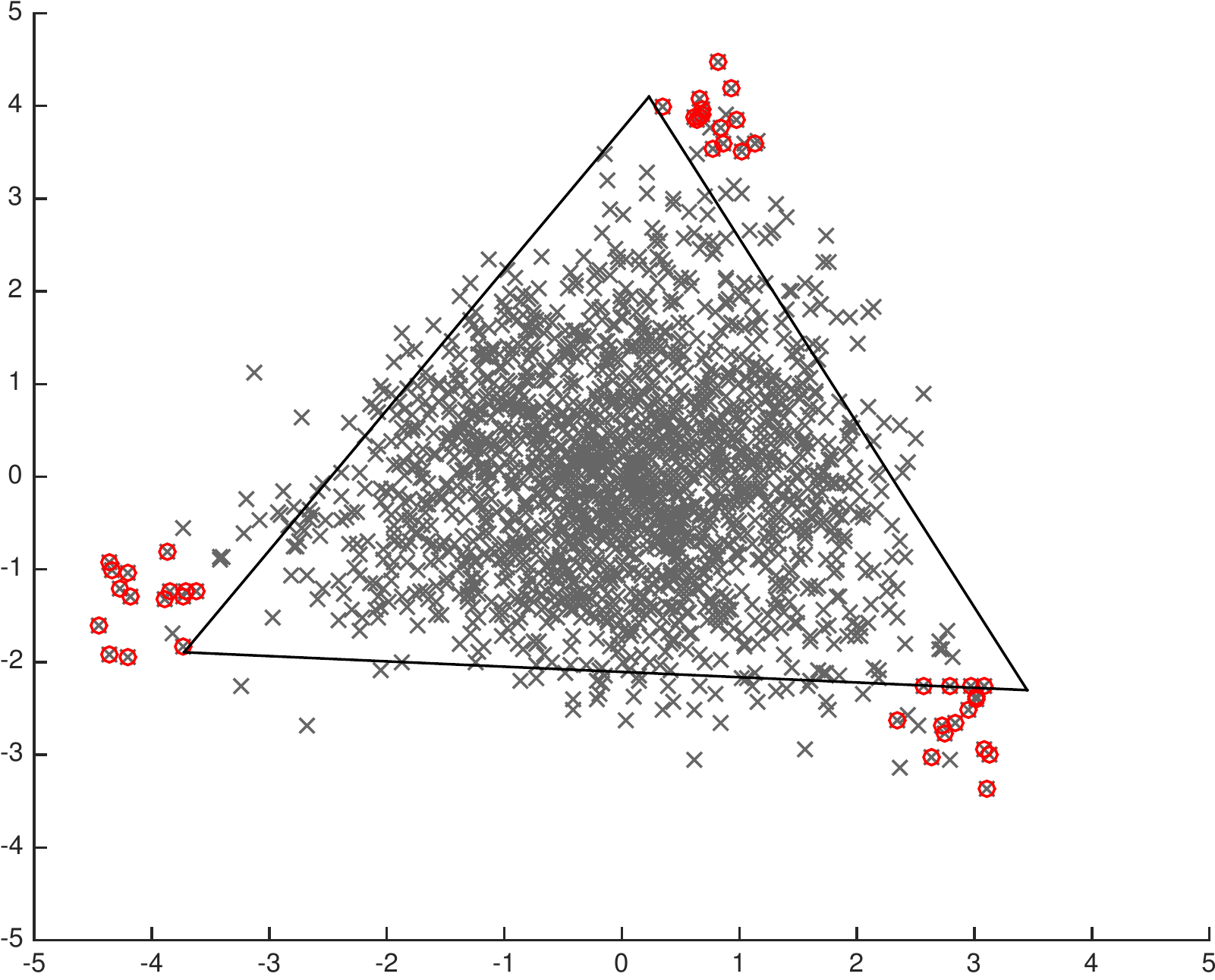}
\includegraphics[width = .4\textwidth, height=.28\textwidth, trim=30 40 10 0, clip=true]{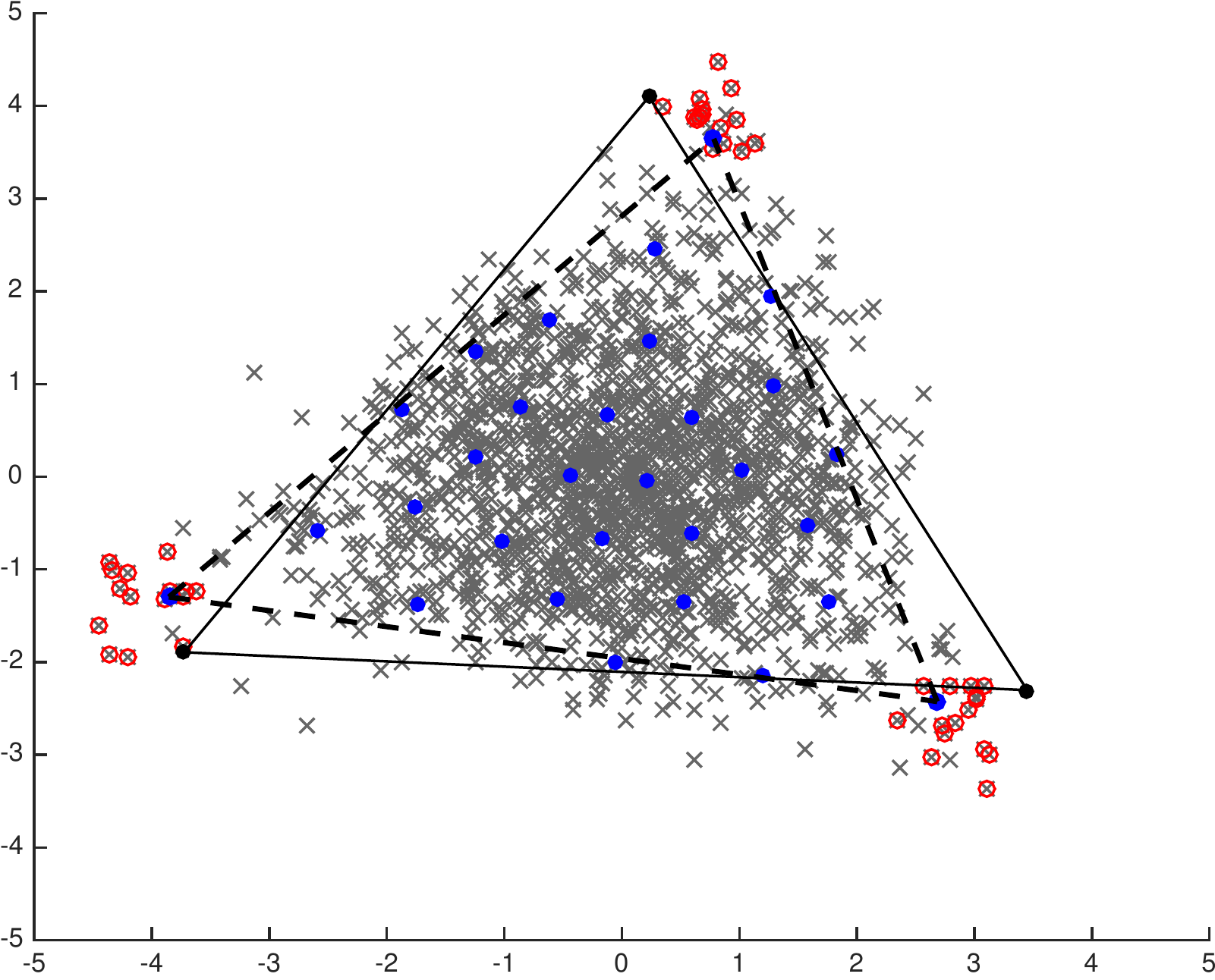}
\caption{The simplex structures in topic models ($K=3$). Top left: Rows of $\tilde{A}V$ and the simplex ${\cal S}_K$ (which lives in $\mathbb{R}^K$). Top right: Rows of $R$ and the simplex ${\cal S}_K^*$ (which lives in $\mathbb{R}^{K-1}$). Bottom left: Rows of $\hat{R}$, where the simplex ${\cal S}_K^*$ has been corrupted by noise. Bottom right: The vertices estimation algorithm, where the blue points are the local centers found by the $k$-means algorithm, and the dashed triangle is the estimated $\hat{\cal S}_K^*$. In all four panels, the anchor words are marked red, where we only see three red points in the top two panels because many anchor words overlap in the noiseless case.}\label{fig:simplex}
\end{figure}

Such a simplex is illustrated in Figure~\ref{fig:simplex} (top left panel). Lemma~\ref{lem1} has some good news and some bad news. 
\begin{itemize}
\item The {\bf good news} is, there is a 
simplex structure which can be used to reconstruct $v_1, v_2, \ldots, v_K$ and so $\tilde{a}_i$: We can first use rows of $\tilde{A}V$, if available, to identify the vertices $v_1,\cdots, v_K$ (e.g., by a convex hull algorithm). Then, each row of $\tilde{A}V$ can be written as the convex combination of $v_1,\cdots, v_K$ with a unique weight vector, and this vector is exactly $\tilde{a}_i$. Having obtained $\tilde{A}=[\tilde{a}_1, \tilde{a}_2, \ldots, \tilde{a}_p]^T$, we can then recover the matrix $A$ by renormalizing each column of $\tilde{A}$ to have a sum of $1$.  
\item The {\bf bad news} is, the simplex structure is {\it associated with the matrix $\tilde{A} V$, not with  $\Xi$} as we had hoped: if we view each row of $\Xi=\diag(h)\cdot \tilde{A}V$ as a point in $\mathbb{R}^K$ in a similar fashion, we do not see such a simplex structure, for each row of $
\Xi$ is the result of scaling the corresponding row  of $\tilde{A} V$, where the scaling factors  vary from one occurrence to the other.  
\end{itemize}
The problem is then how to fix the bad news while keeping the good news. 

Here is our proposal. Define a $p \times (K-1)$ matrix $R = [R_1, R_2,   \ldots, R_{K-1}]$  by taking the entrywise ratios of the $k$-th singular vector and the first singular vector, for $2\leq k\leq K$: 
\beq \label{defineR}
R_k(i) = \xi_{k+1}(i) / \xi_1(i), \qquad 1 \leq k \leq K-1, \;\; 1 \leq i \leq p. 
\eeq
By Perron's theorem \cite{HornJohnson}, $\xi_1$ has all positive entries, so $R$ is well-defined. Since each row of $\Xi$ is the result of scaling the corresponding row  of $\tilde{A} V$, we hope that the operation of ``taking the entrywise ratios" will remove such undesirable scaling effects. Below, we demonstrate that, if we view rows of the matrix $R$ as points in $\mathbb{R}^{K-1}$, there is a low-dimensional simplex that can be used to reconstruct $A$. 

Recall the definition of $V$ in \eqref{defineV} and write $V=[V_1,\cdots, V_K]$. 
We introduce a $K\times (K-1)$  matrix $V^* = [V_1^*, V_2^*, \ldots, V_{K-1}^*]$ by taking the entrywise ratios of the $k$-th column of $V$ and its first column, for $2\leq k\leq K$: 
\beq \label{defineVstar}
V_k^*(j) = V_{k+1}(j)/V_1(j), \qquad 1 \leq k \leq K-1,\;\; 1 \leq j \leq K.
\eeq
By definitions~\eqref{defineR}-\eqref{defineVstar}, $[1_p, R]=[\mathrm{diag}(\xi_1)]^{-1}\Xi$, and $ [1_K, V^*] = [\mathrm{diag}(V_1)]^{-1}V$, where $1_m$ is the $m$-dimensional vector of $1$'s. Combining these with \eqref{defineV}, we find that  
\beq \label{definePi}
R = \Pi \cdot V^*, \qquad \mbox{where}\quad \Pi \equiv [\mathrm{diag}(\xi_1)]^{-1} A \cdot \mathrm{diag}(V_1)\in\mathbb{R}^{p,K}. 
\eeq
Therefore, each row of $R$ is a convex combination of the $K$ rows of $V^*$, with the weight vector being the $i$-th row of $\Pi$ (it can be verified that each row of $\Pi$ is indeed a weight vector; see Section~\ref{sec:proof}). 

The above calculations give the key observation of this paper. Letting $v_1^*,\cdots, v_K^*\in\mathbb{R}^{K-1}$ be the rows of the matrix $V^*$, we define ${\cal S}_K^*$ as the simplex in $\mathbb{R}^{K-1}$ with $v_1^*, v_2^*, \ldots, v_K^*$ being the vertices. The following lemma is a direct result of \eqref{definePi}.  

\begin{lem}[Key observation] \label{lem:key}
Let $r_1,\cdots,r_p\in\mathbb{R}^{K-1}$ be the rows of the matrix $R$.  
If word $i$ is an anchor word, then $r_i$  coincides with one vertex of ${\cal S}_K^*$. If word $i$ is not an anchor word, then $r_i$ falls into the interior of ${\cal S}_K^*$ (or the interior of an edge/face of ${\cal S}^*_K$); moreover,  $r_i$ is also a convex combination of 
 $v_1^*, v_2^*, \ldots, v_K^*$ with $\pi_i$ being the weight vector, where $\pi_i$ is the $i$-th row of the matrix 
$\Pi = [\diag(\xi_1)]^{-1}   A  \cdot    \diag(V_1)$, $1 \leq i\leq p$. 
 \end{lem} 

Such a simplex is illustrated in Figure~\ref{fig:simplex} (top right panel).
We are now ready to construct $A$ by $\Xi$, with the following steps: 
\begin{enumerate}
\item[(i)]  Obtain $R$ using \eqref{defineR}. 
\item[(ii)] Viewing the rows of $R$ as points in $\mathbb{R}^{K-1}$, use the simplex structure to identify the vertices $v_1^*, v_2^*, \ldots, v_K^*$ (say, by a convex hull algorithm).
\item[(iii)] Use $r_j$ and $v_1^*, v_2^*, \ldots, v_K^*$ to obtain $\pi_j$ (and so $\Pi$)  through the relationship of $r_j = \sum_{\ell = 1}^K \pi_j(\ell) v_{\ell}^*$.
\item[(iv)] Obtain the matrix   
$A  \cdot  \diag(V_1)$ using $(\xi_1, \Pi)$ with the relationship of $\Pi = [\diag(\xi_1)]^{-1}  \cdot    A   \cdot     \diag(V_1)$.
\item[(v)] Normalize the matrix $A \cdot  \diag(V_1)$ column-wise so that each column has a sum of $1$. The resultant matrix is seen to be $A$. 
\end{enumerate}


\section{A practical SVD-based method}  \label{sec:method}
In the real case where $\hat{\Xi}$, instead of $\Xi$, is available, 
most of the above steps are easy to generalize, except for (ii): we observe a noise corrupted version of ${\cal S}^*_K$ (see Figure~\ref{fig:simplex}, bottom panels), and we need to estimate $v_1^*, \cdots, v_K^*$ from the noisy data. Below, we first describe our method without details of estimating the vertices, and then introduce a practical algorithm for vertices estimation. 

{\bf Our method.} Input: $D, K$. Output: $\hat{A}$. Tuning parameter: $s$ (max number of words in each topic). 

\begin{enumerate}
\item Apply SVD to $D$, and let $\hat{\xi}_1,\cdots,\hat{\xi}_K$ be the first $K$ left singular vectors. Compute the matrix $\hat{R}$ by $\hat{R}(j,k) = {\cal T}( \hat{\xi}_{k+1}(j)/\hat{\xi}_1(j), \log(n\vee p) )$, for  $1\leq j\leq p$, $1\leq k\leq K-1$, 
where ${\cal T}$ is a truncation function such that ${\cal T}(x,a)=\mathrm{sign}(x)\cdot \min\{|x|, a\}$. 
Let $\hat{r}_1,\cdots,\hat{r}_p$ be the rows of $\hat{R}$. 

\item Apply a vertices estimation algorithm to $\hat{r}_1,\cdots, \hat{r}_p$. Let $\hat{v}^*_1,\cdots,\hat{v}^*_K$ be the output of the algorithm. Write $\hat{V}^*=[\hat{v}_1^*, \cdots, \hat{v}^*_K]^T$. 
\item For each $1\leq j\leq p$, obtain $\tilde{\pi}_j$ by  $\tilde{\pi}^T_j =[ \hat{r}_j^T, 1][\hat{V}^*, {\bf 1}_K]^{-1}$, where ${\bf 1}_K$ is the $K$-dimensional vector all the entries of which are $1$. We truncate the negative entries in $\tilde{\pi}_j$ and renormalize it: let $\hat{\pi}^*_j(k)=\max\{\tilde{\pi}_j(k), 0\}$, $1\leq k\leq K$, and $\hat{\pi}_j=\hat{\pi}_j^*/\|\hat{\pi}_j^*\|_1$. Write $\hat{\Pi}=[\hat{\pi}_1,\cdots,\hat{\pi}_p]^T$.  
\item Obtain the matrix $\hat{A}^*=\hat{\Pi}\cdot\mathrm{diag}(\hat{\xi}_1)$. Write $\hat{A}^*=[\hat{A}_1^*, \cdots, \hat{A}^*_K]$. 
\item For each $1\leq k\leq K$, keep the $s$ largest entries of $\hat{A}^*_k$ and set all other entries to $0$; denote the resultant vector by $\hat{A}_k^{**}$. Let $\hat{A}_k=\hat{A}^{**}_k/\|\hat{A}^{**}_k\|_1$. Output $\hat{A}=[\hat{A}_1,\cdots,\hat{A}_K]$.  
\end{enumerate}

{\bf Remark}. Our method largely preserves the ``non-negativeness'' feature of the problem, although the original SVD does not. 
In Step 3, as long as $\hat{r}_i$ is within the estimated simplex (the majority of them do; see Figure~\ref{fig:simplex}), $\tilde{\pi}_i$ is a valid weight vector, so there is no need to truncate. 

We then introduce a vertices estimation algorithm. It is motivated by the following observation. {\it In the idealized case, for all the anchor words of one topic, their associated $r_i$'s fall onto the same point --- one vertex $v^*_k$. In the noisy case, the corresponding $\hat{r}_i$'s form a local data cluster around $v^*_k$.} Therefore, our proposal is to first identify a few local data cluster centers, say, by a classical $k$-means algorithm assuming $>K$ clusters, and then use the simplex geometry to estimate the vertices from these local centers. 

{\bf Algorithm for vertices estimation.} Input: $\hat{r}_1,\cdots,\hat{r}_p$. Output: $\hat{v}^*_1,\cdots,\hat{v}^*_K$. Tuning parameter: $m$ (number of clusters in $k$-means). 
\begin{enumerate}
\item[2a.] Apply the classical $k$-means algorithm to $\hat{r}_1,\cdots,\hat{r}_p$, with the number of clusters set to be $m$. Let $\hat{\theta}_1,\cdots,\hat{\theta}_m$ be the cluster centers given by $k$-means.  
\item[2b.] Search among $1\leq j_1< j_2<\cdots< j_K\leq m$ such that $\hat{\theta}_{j_1},\cdots,\hat{\theta}_{j_K}$ are affinely independent, and find $(\hat{j}_1,\cdots,\hat{j}_K)$ that minimizes 
\[
\max_{1\leq j\leq m} \big\{\mathrm{distance}\big(\hat{\theta}_j, \; {\cal S}(\hat{\theta}_{j_1},\cdots,\hat{\theta}_{j_K})  \big)\big\},\footnote{For any point $v$ and set ${\cal S}$ in $\mathbb{R}^{K-1}$,  $\mathrm{distance}(v,{\cal S})$ denotes the distance from $v$ to ${\cal S}$. This distance is zero if $v$ is in ${\cal S}$. For ${\cal S}$ being a simplex,  $\mathrm{distance}(v,{\cal S})$ can be conveniently computed via a quadratic programming with $K$ variables.}
\]
where ${\cal S}(\hat{\theta}_{j_1},\cdots,\hat{\theta}_{j_K})$ is the simplex with vertices $\hat{\theta}_{j_1},\cdots,\hat{\theta}_{j_K}$. Output $\hat{v}^*_k=\hat{\theta}_{\hat{j}_k}$, $1\leq k\leq K$. If no such $(\hat{j}_1,\cdots,\hat{j}_K)$ exist, output $\hat{v}^*_k=e_k$ (the standard basis vector), $1\leq k\leq K$. 
\end{enumerate}

The algorithm is illustrated in Figure~\ref{fig:simplex} (bottom right panel). The 2a step can be viewed as {\it complexity reduction}, where we ``sketch'' the data cloud by only $m$ local centers and discard all the original data points. In the 2b step, we try to find the ``best-fit'' simplex from the $m$ centers, by a combinatorial search. 


The method has two tuning parameters $(s, m)$. Noting that $s$ is the maximum number of words allowed in each topic, we usually set $s=p$. For $m$, our numerical experiments suggest a choice of $m$ being $10\sim 20$ times of $K$ works well; see Section~\ref{sec:numeric}. 

We now discuss the computational complexity of our method, which is dominated by the SVD step and the vertices estimation step. While the complexity of full SVD is $O(np\cdot \min\{n,p\})$, the complexity of our SVD step is only $O(Knp)$ for we only need the first $K$ left singular vectors \cite{halko2011finding}. The complexity of the vertices estimation step is $O(K!)$, independent of $(n,p)$. 
In practice, the vertices estimation step is reasonably fast for moderately large $K$ (about $2$ minutes for $K = 20$). There is also much space for improving the computation of our method. For example, 
for very large $n$, we can use randomized algorithms \cite{frieze2004fast} for SVD. For relatively large $K$ (e.g., $K=50$), we can significantly speed up the vertices estimation step by replacing the exhaustive search by a greedy search; see below. 

In the 2b step, it is unnecessary to search all $K$-tuples of the local centers, noting that, the local centers that are located deeply in the middle of the data cloud contribute very little to estimating the vertices. Hence, to speed up the algorithm, we first use a greedy search to remove most local centers but only keep $K_0$ of them (say, $K_0=2K$), and then apply the 2b step. 

\begin{enumerate}
\item[2b'.] Find the two local centers $\hat{\theta}_{j_1}$ and $\hat{\theta}_{j_2}$ such that the distance between them is the maximum over all pairs of local centers. For $k=2,3, \cdots, K_0$, let $\hat{\theta}_{j_k}$ be the local center such that its distance to $\bar{\theta}_{(k-1)}\equiv \frac{1}{k-1}\sum_{\ell<k}\hat{\theta}_{j_\ell}$ is the maximum over all $\hat{\theta}_j\notin \{\hat{\theta}_{j_\ell}\}_{\ell=1}^{k-1}$. We then apply the 2b step to $\hat{\theta}_{j_1}, \cdots, \hat{\theta}_{j_{K_0}}$ only.  
\end{enumerate}

\section{Asymptotic analysis}  \label{sec:theory}
Without loss of generality, we assume all the documents have the same length $N$ (i.e., consisting of $N$ words). In the pLSI model, the documents are generated independently of each other, and for document $i$, the $N$ words are $iid$ drawn from the $p$ vocabulary words using the $i$-th column of $D_0$. Write $Z=[Z_1,\cdots, Z_n]$ and $D_0=[d^0_{1}, \cdots, d^0_{n}]$. Then, $Z_i$'s are independent of each other, and 
\beq \label{Mod2}
Z_i \overset{(d)}{=} N^{-1}(X_i - E[X_i]), \qquad X_i \sim \mathrm{Multinomial}(N, d_i^0), \quad 1\leq i\leq n. 
\eeq
We note that \eqref{Mod2} comes directly from the pLSI model. 

By \eqref{Mod1}, $D_0=AW$; for this factorization to be unique, we need an identifiability condition. We use the one in \cite{donoho2003does} (see \cite{ding2013necessary} for other identifiability conditions): each topic $k$ has at least one anchor word and at least one pure document (i.e., $W_i(k)=1$). Without loss of generality, we assume the sum of each row of $A$ is nonzero (otherwise, with probability $1$, this word appears in none of the documents).  

Define the ``topic-topic correlation'' matrix $\Sigma_A= A^TA \in\mathbb{R}^{K,K}$ and the ``topic-topic concurrence'' matrix $\Sigma_W =  n^{-1}WW^T\in\mathbb{R}^{K,K}$. 
We use an asymptotic framework with $n,p$ tending to infinity, where $K$ is fixed, and for a fixed non-negative matrix $\Sigma_0$ and a sequence $\{\beta_p\}_{p=1}^{\infty}$, 
\beq \label{cond-A}
\lim_{p\to\infty} \beta_p^{-1}\Sigma_A = \Sigma_0, \qquad \mbox{$\Sigma_0$ is non-singular and irreducible}. 
\eeq
Since the row sum of $A$ is $1$, $\lambda_{\max}(\Sigma_A)$ is bounded. So this sequence $\beta_p$ should be bounded, and it is allowed to tend to $0$ as $p\to\infty$. 
For $\Sigma_W$, we assume for constants $c_1, c_2>0$, 
\beq \label{cond-W}
\lambda_{\min}(\Sigma_W)\geq c_1, \qquad \lambda_1(\tilde{\Sigma}_W)-\lambda_2(\tilde{\Sigma}_W)\geq c_2, \qquad \tilde{\Sigma}_W\equiv \Sigma_0^{1/2}\Sigma_W\Sigma_0^{1/2}, 
\eeq
where for any matrix $M$ and an integer $k\geq 1$, $\lambda_k(M)$ denotes the $k$-th largest eigenvalue of $M$. Compared with literature, we need neither the ``topic imbalance'' condition in \cite{arora2013practical} (excluding pure documents) nor the ``dominant topic'' condition in \cite{bansal2014provable} (excluding highly mixing documents). 

To state our results, we need some notations. Recall that $A_k$ is the $k$-th column of $A$ and $a^T_j$ is the $j$-th row of $A$, and $h(j)=\|a_j\|_1$ and $\tilde{a}_j=[h(j)]^{-1} a_j$.

\begin{defin}\label{def:notations}
Let $h_{\min}=\min_{1\leq j\leq p}\|a_j\|_1$ and $h_{\max}=\max_{1\leq j\leq p}\|a_j\|_1$, where $\|a_j\|_1$ describes the ``overall frequency'' of word $j$ in all topics. Let $s_p = \max_{1\leq k\leq K}\{\|A_k\|_0\}$ be the maximum number of words in one topic. Let $\gamma_n= \max_{1\leq j\leq p}\{\sum_{i=1}^n D_0(j,i)\}$, where $\sum_{i=1}^n D_0(j,i)$ is the ``overall appearance'' of word $j$ in all documents. 
\end{defin}


\begin{lem}[The simplex]  \label{lem:simplex}
Under the pLSI model \eqref{Mod1} and \eqref{Mod2},   
as $n,p,N\to\infty$, suppose \eqref{cond-A}-\eqref{cond-W} hold. There are constants $C_1, C_2>1$ such that the distance between any two vertices of ${\cal S}^*_K$ is bounded above by $C_1$ and below by $C_1^{-1}$, and the volume of ${\cal S}^*_K$ is bounded above by $C_2$ and below by $C_2^{-1}$.  
\end{lem}


\begin{lem}[Noise perturbatiton] \label{lem:noise}
Under the pLSI model \eqref{Mod1} and \eqref{Mod2}, as $n,p,N\to\infty$,  suppose \eqref{cond-A}-\eqref{cond-W} hold. With probability $1-o((n\vee p)^{-2})$, there exists a $(K-1)\times (K-1)$ orthogonal matrix $O$ such that 
\[
\sum_{j=1}^p \|\hat{r}_j-Or_j\|^2 \leq C \epsilon_n^2, \qquad \epsilon_n\equiv \frac{1}{h_{\min}} \left( \frac{\sqrt{s_p h_{\max}}}{\sqrt{Nn}} + \frac{(\gamma_n\vee 1)}{N n\sqrt{\beta_p}}\right)\log^2(n\vee p). 
\]
\end{lem} 

{\bf Remark}. The proof of Lemma~\ref{lem:noise} requires delicate matrix perturbation analysis (e.g., the sine-theta theorem \cite{sin-theta}) and random matrix theory (e.g., the matrix Bernstein inequality \cite{tropp2012user}). Especially, we note that the entries of the noise matrix $Z$ are not independent, which poses technical difficulty. 

From Lemmas~\ref{lem:simplex}-\ref{lem:noise}, it is hopeful to estimate the vertices up to an $\ell_2$-error of $O(p^{-1/2}\epsilon_n)$. 
In the next theorem, we assume such an vertices estimation algorithm is available, and evaluate the performance of our method. 

\begin{thm}[Error rate]\label{thm:rate}
Under the pLSI model \eqref{Mod1} and \eqref{Mod2}, as $n,p,N\to\infty$, suppose \eqref{cond-A}-\eqref{cond-W} hold, and let $\epsilon_n$ be the same as that in Lemma~\ref{lem:noise}. Suppose that the tuning parameter $s$ satisfies  $s_p\leq s\leq Cs_p$ with $s_p$ as in Definition~\ref{def:notations}, and that with probability $1-o((n\vee p)^{-2})$, the vertices estimation algorithm satisfies $\max_{1\leq k\leq K}\{\|\hat{v}^*_k- v^*_{\omega(k)}\|\}\leq Cp^{-1/2}\epsilon_n$ for  a permutation $\omega(\cdot)$ on $1,\cdots, K$. With probability $1-o((n\vee p)^{-2})$, 
\[
\sum_{k=1}^K \|\hat{A}_k - A_{\omega(k)}\|_1 \leq C\frac{h_{\max}}{h_{\min}} \left(\frac{s_p\sqrt{h_{\max}}}{\sqrt{Nn\beta_p}} + \frac{(\gamma_n\vee 1)\sqrt{s_p}}{N n\beta_p}\right)\log^2(n\vee p).
\] 
\end{thm}

In a special case where all the entries of $A$ are in the same order (so $\beta_p\asymp h_{\max}\asymp 1/p$, $s_p=p$ and $\gamma_n\leq nh_{\max}$), the $\ell_1$-estimation error is $O\left( (Nn)^{-1/2}p + (Nn)^{-1}p^{3/2} + N^{-1}p^{1/2}\right)$. For many data sets where the length of document, $N$, is appropriately large, the result suggests that our method works well. Furthermore, in the case $n\gg N$, if we consider an idealized model where the entries of $Z$ are independent rather than being generated from \eqref{Mod2}, we conjecture that the term $N^{-1}p^{1/2}$ in the error rate can be improved to $(nN)^{-1/2}p^{1/2}$. 

{\bf Remark}. The error rate is not the same as that in \cite{arora2013practical} because the settings are very different. For example, we assume the columns of $W$ are non-random rather than being $iid$ drawn from a distribution, so our setting has a lot more nuisance parameters as $n$ increases. 

We now investigate the vertices estimation algorithm in Section~\ref{sec:method}. It requires additional conditions. For $1\leq k\leq K$, let $\mathcal{N}_k=\{1\leq j\leq p: \pi_j(k)=1\}$ be the set of anchor words of topic $k$, and let ${\cal M}$ be the set of non-anchor words. Recall that $\tilde{a}_j=\|a_j\|_1^{-1}a_j$ is a weight vector describing the relative frequency of word $j$ across topics. 
We need a mild ``concentration'' on the $\tilde{a}_j$'s of non-anchor words: for a fixed integer $L_0\geq 1$, there are weight vectors $b_1^*,\cdots, b_{L_0}^*\in\mathbb{R}^{K}$ satisfying ($e_k$ is the $k$-th standard base vector)
\beq \label{cond-alg1}
\min\big\{\min_{1\leq i\neq j\leq L_0}\|b_i^*-b_j^*\|, \; \min_{1\leq k\leq K, 1\leq i\leq L_0}\|b_i^*- e_k\|\big\}\geq c_3, 
\eeq 
and a partition of ${\cal M}$, ${\cal M}={\cal M}_1\cup{\cal M}_2\cup\cdots\cup{\cal M}_{L_0}$, such that 
\beq \label{cond-alg2}
\min_{1\leq \ell\leq L_0}\{|{\cal M}_\ell|/|{\cal M}|\}\geq c_4, \qquad  \alpha_p\equiv \max_{1\leq\ell\leq L_0, j\in{\cal M}_\ell}\{\|\tilde{a}_j - b_k^*\|\}\to 0. 
\eeq
Furthermore, depending on $\alpha_p$ (the degree of ``concentration'' of the non-anchor words) and $\epsilon_n$ (the noise level), we need enough number of anchor words for each topic: 
\beq  \label{cond-alg3}
\min_{1\leq k\leq K}\{|{\cal N}_k|\}\geq (|{\cal M}|\cdot \alpha^2_p + \epsilon_n^2) \cdot \log(n\vee p). 
\eeq

\begin{thm}[Vertices estimation] \label{thm:VHalg}
Under the pLSI model \eqref{Mod1} and \eqref{Mod2}, as $n,p,N\to\infty$, suppose \eqref{cond-A}-\eqref{cond-W} hold, and additionally \eqref{cond-alg1}-\eqref{cond-alg3} hold, and let $\epsilon_n$ be the same as that in Lemma~\ref{lem:noise}.  With probability $1-o((n\vee p)^{-2})$, for a proper $m$ in the vertices estimation algorithm, there is a permutation $\omega(\cdot)$ on $1,\cdots, K$ such that $\max_{1\leq k\leq K}\|\hat{v}^*_k- v^*_{\omega(k)}\|\leq C p^{-1/2}\epsilon_n$, and so $\sum_{k=1}^K \|\hat{A}_k - A_{\omega(k)}\|_1$ has the same upper bound as that in Theorem~\ref{thm:rate}. 
\end{thm}



\section{Numerical experiments} \label{sec:numeric}



\subsection{Simulations}

We compare our method with the standard LDA on simulated data. In our method, we set $s=p$ and $m=10K$, where $K$ is the number of topics; also, in the vertices estimation algorithm, we use 2b', instead of 2b, with $K_0=\lceil 5K/4\rceil$. To implement the standard LDA, we use the R package {\it lda} with the two Dirichlet hyper-parameters both set to be $0.1$. For both methods, we evaluate the $\ell_1$-estimation error $\min_{\omega(\cdot)}\{\max_{1\leq k\leq K} \|\hat{A}_k-A_{\omega(k)}\|\}$, where the minimum is over all permutations on $1,\cdots, K$.

Fixing $(K, n,p,N, a_0, p_0)$, we generate the data as follows. (i) We first generate $A$. Let the first $Kp_0$ words be anchor words, $p_0$ for each topic, with the corresponding entry of $A$ being $1.5/p$. For each topic $k$, we generate the remaining $(p-Kp_0)$ entries independently from $p^{-1} U(0,1)$, where $U(0,1)$ is the uniform distribution over $(0,1)$. Finally, we renormalize each column of $A$ to have a sum of $1$. (ii) We then generate $W$. Set the first $n\alpha_0$ documents to be pure documents, $(n\alpha_0)/K$ for each topic. For the remaining $n(1-\alpha_0)$ columns, we generate them independently, where entries of $W_i$ are first $iid$ drawn from $U(0,1)$ and then the column is normalized to have a sum $1$. (iii) We last generate $Z$ according to model \eqref{Mod2}. In all the experiments, we fix $p=2000$ and $p_0=20$ (so each topic has $1\%$ of anchor words). 

{\it Experiment 1: choice of tuning parameters}. We fix $K=6$, $N=2000$, $n=500$ and $a_0=0.2$, and implement our method for different choices of $m$. The $\ell_1$ estimation error (averaged over $50$ repetitions) are displayed below:
\begin{center}\begin{tabular}{c|cccccc}
\hline
$m$ & 12 & 24 & 36 & 48 & 60 & 84\\
\hline
error & .190 & .188 & .187 & .189 & .186 & .187 \\
\hline
\end{tabular}\end{center}
It suggests that the performance of our method is similar for a range of $m$.  

\begin{figure}[t]
\centering
\includegraphics[width = .32\textwidth, height=.25\textwidth]{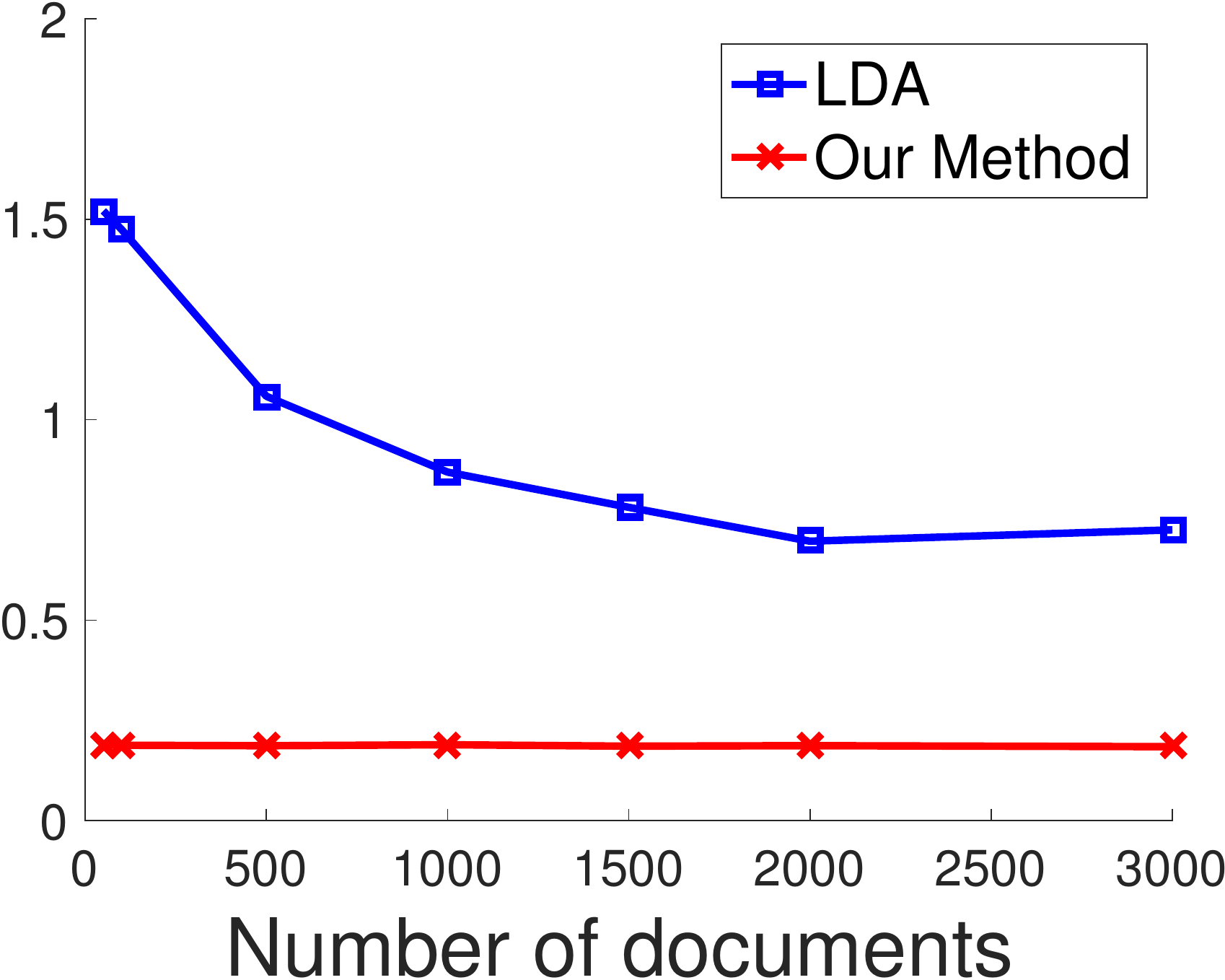}
\includegraphics[width = .32\textwidth, height=.25\textwidth]{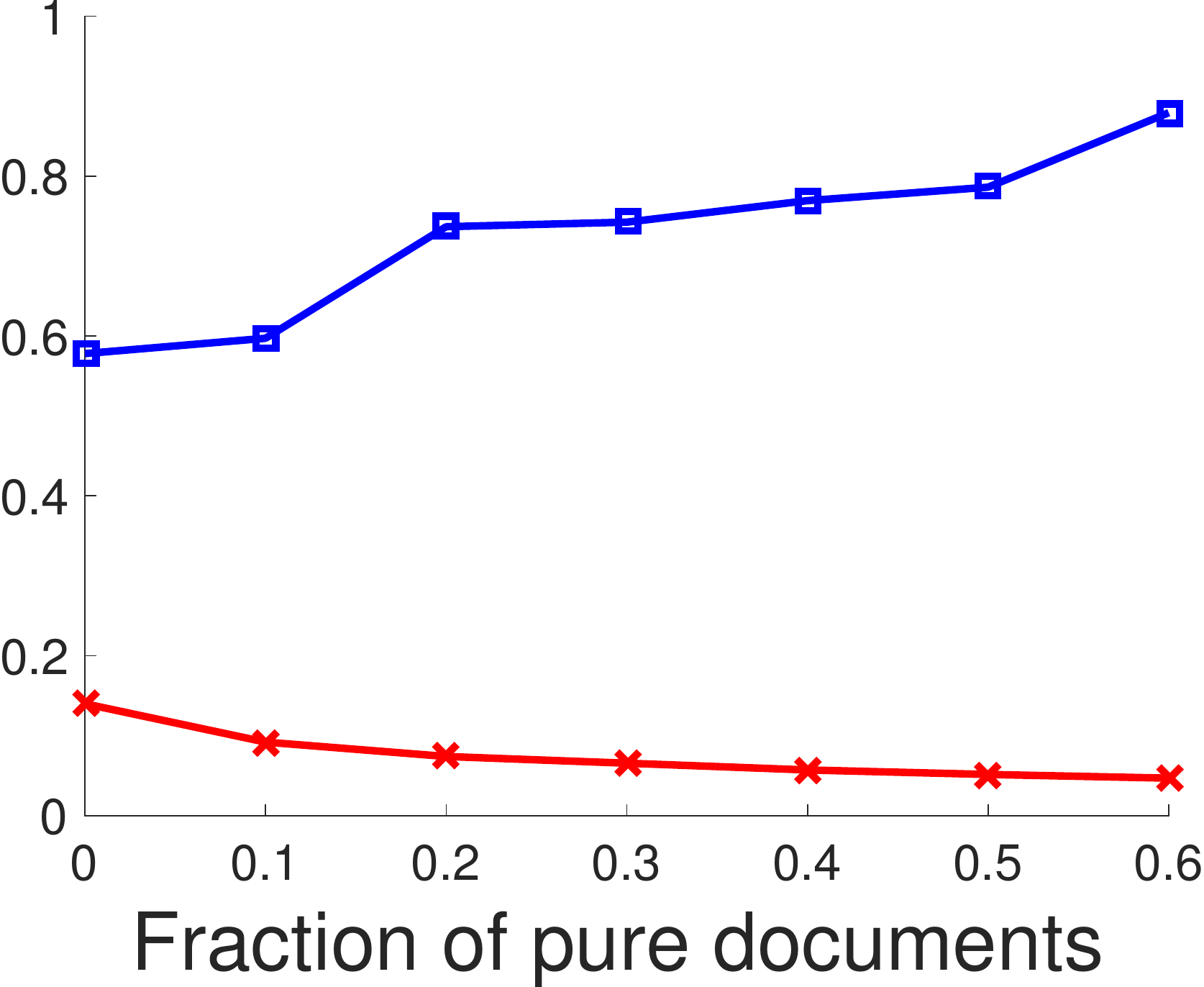}
\includegraphics[width = .32\textwidth, height=.25\textwidth]{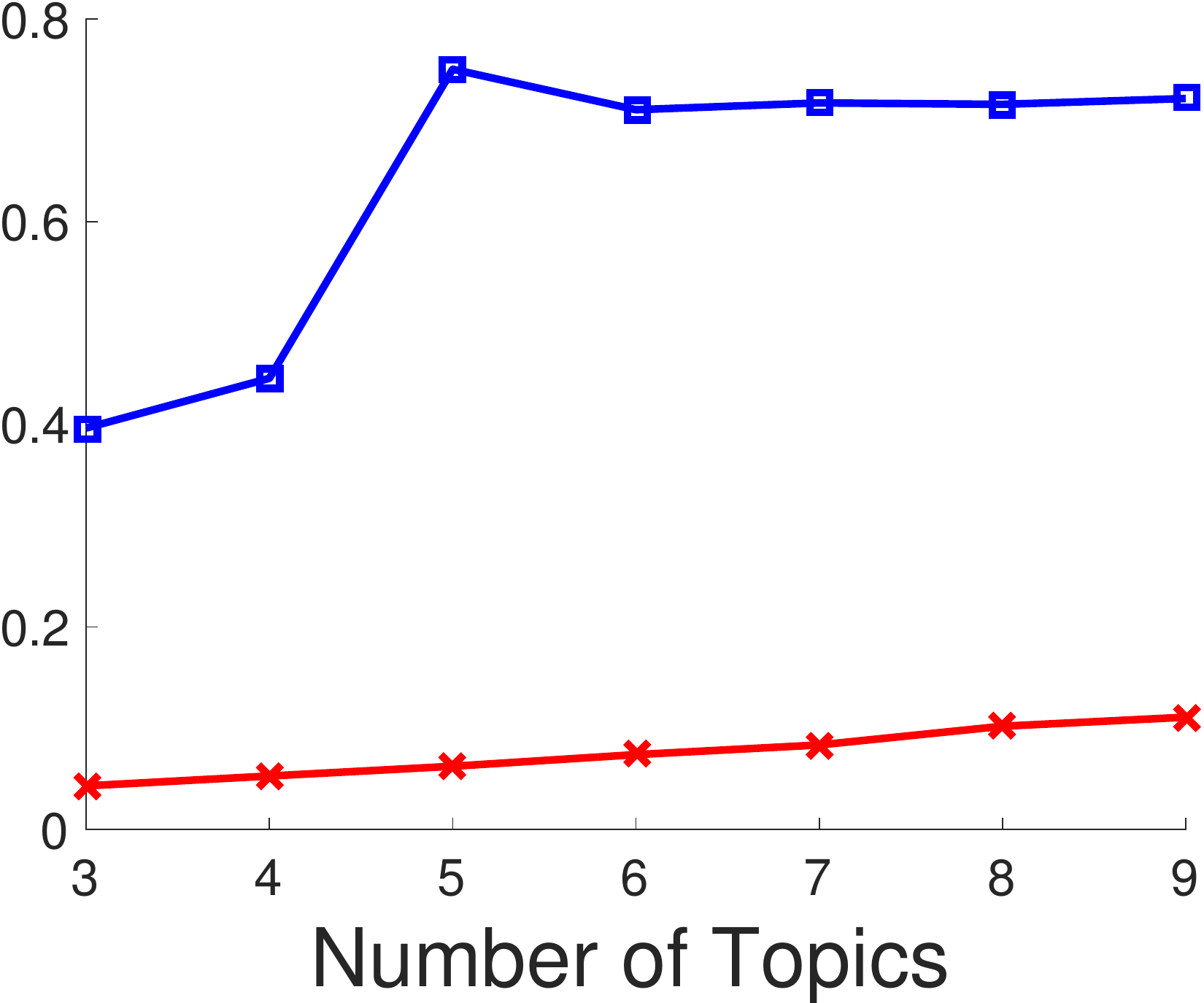}
\caption{Comparison of our method and LDA on simulated data. Y-axis: the $\ell_1$ estimation error; x-axis:  the number of documents $n$ (left), the fraction of pure documents $a_0$ (middle), and the number of topics $K$ (right).}\label{fig:errors}
\end{figure}

{\it Experiment 2: total number of documents}. We fix $K=6$, $N=2000$, $a_0=0.2$ and let $n$ take different values in $\{50, 100, 500, 1000, 1500, 2000, 3000\}$. The results are in Figure~\ref{fig:errors} (left panel). It suggests that the performance of our method improves quickly as $n$ increases, especially when $n$ is relatively small. Also, our method uniformly outperforms LDA. 


{\it Experiment 3: fraction of pure documents}. We fix $K=6$, $N=2000$, $n=3000$ and let $a_0$ increase from $0\%$ to $60\%$ (so the fraction of pure documents increases). The results are in Figure~\ref{fig:errors} (middle panel). It suggests that the performance of our methods improves as there are more pure documents, and our method outperforms LDA. 

{\it Experiment 4: number of topics}. We fix $N=2000$, $n=3000$, $a_0=0.2$, and let $K$ take different values in $\{3, 4, \cdots, 9\}$. The results are in Figure~\ref{fig:errors} (right panel). The performance of our method deteriorates slightly as $K$ increases, but it is always better than LDA.

\begin{figure}[t]
\centering
\includegraphics[width = .3\textwidth, height =.22\textwidth]{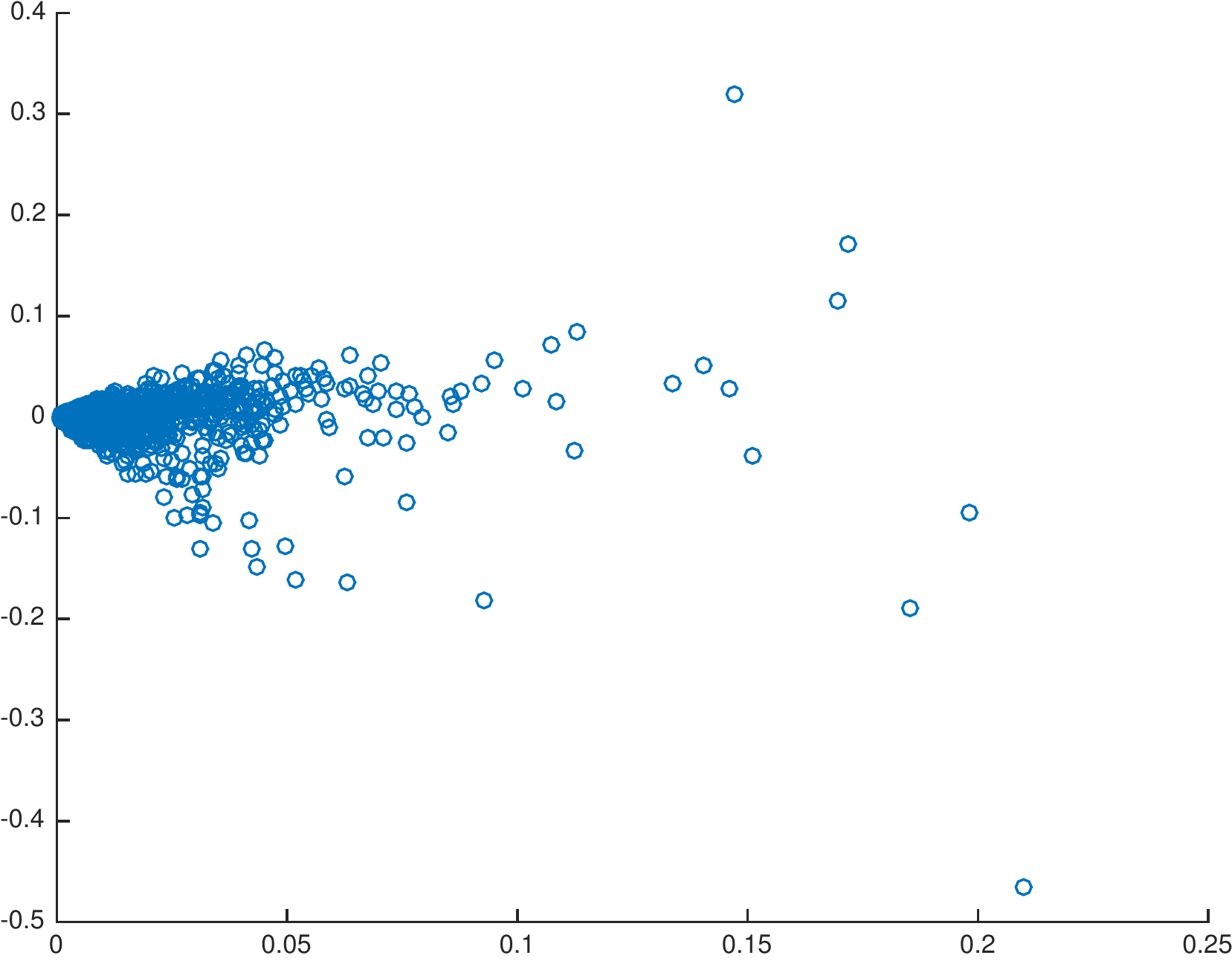}
\includegraphics[width = .3\textwidth, height =.22\textwidth]{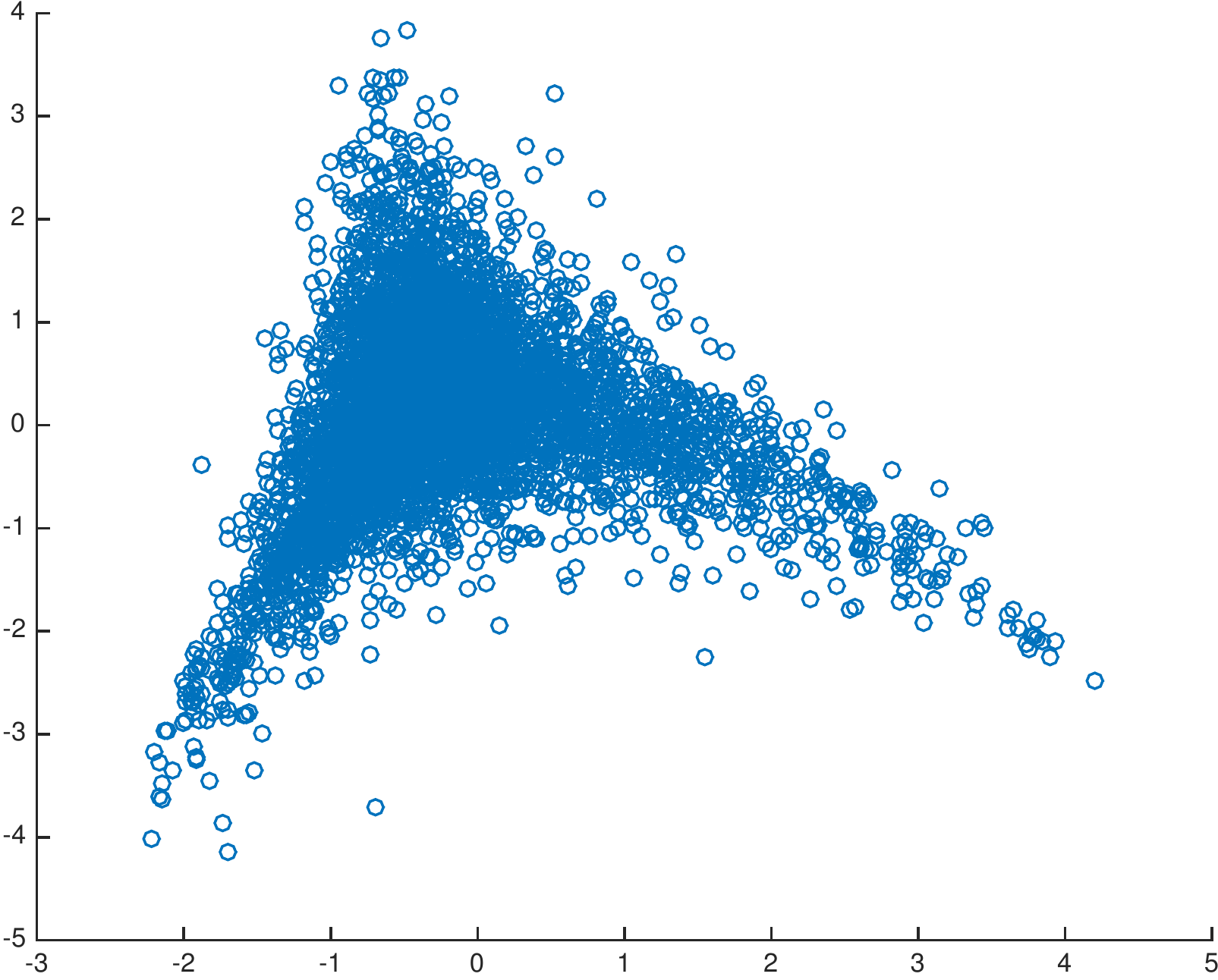} 
\caption{Associated Press (AP) data. The left and right panels contain the scatter plots of the first two columns of $\hat{\Xi}$ and those of $\hat{R}$, respectively. }\label{fig:eigenvec}
\end{figure}

\subsection{Real data}
We take the {\it Associated Press} (AP) data set \cite{harman1overview}, consisting of $2246$ documents and a vocabulary of $10473$ words. As preprocessing, we first remove a list of $40$ stop words, an then keep only $5000$ most frequent words in the vocabulary; also, we remove documents whose length is small, and keep only $95\%$ of the documents. 

In Figure~\ref{fig:eigenvec} (left panel), we plot the first two columns of $\hat{\Xi}$, where each row is viewed as a point in $\mathbb{R}^2$. As expected, the row-wise scaling effect is so strong that no simplex is revealed. In Figure~\ref{fig:eigenvec} (right panel), we plot the first two columns of $\hat{R}$ --- the matrix of eigen-ratios. It is seen that the row-wise scaling effect has been removed, and the data cloud suggests a clear triangle (a simplex with $3$ vertices). The plot also suggests that $K=3$. We apply our methods to $K=3$ and set the tuning parameters as $s=p$ and $m=30$. For each topic, we output the top $20$ words that are most ``anchor-like'' words of this topic (i.e., those words whose $\hat{r}_j$ are closest to the corresponding vertex of that topic); the results are in Table~\ref{tb:AP}. From these words, the three topics seem to be nicely interpreted as ``Crime'', ``Politics'' and ``Finance''. 

\begin{table}
\caption{Estimated topics in the Associated Press data.} \label{tb:AP}
\scalebox{.9}
{\begin{tabular}{ll}
\hline
``Crime'' & {\it police, sikh, dhaka, hindus, shootings, dog, injury, gunfire, bangladesh, gunshot, neck, }\\
& {\it warmus, gunman, wounding, tunnel, searched, gang, blaze, extremists, policemen}\\
\hline
``Politics'' & {\it lithuania, ussoviet, longrange, resolutions, eastwest, boris, ratification, treaty, gorbachev, }\\
& {\it mikhail, norway, gorbachevs, shevardnadze, sakharov, soviet, sununu, yeltsin, cambodia,}\\
& {\it emigration, soviets}\\
\hline
``Finance'' & {\it index, shares, composite, industrials, nyses, exchangelisted, nikkei, gainers, lsqb, outnumbered,}\\
& {\it losers, volume, rsqb, unchanged, traded, points, share, stocks, yen, exchange}\\
\hline
\end{tabular}}
\end{table}


\section{Discussions}  \label{sec:discuss}

The main innovation of this paper is the discovery of a transparent connection between the output of SVD and the targeting topic matrix, in terms of a low-dimensional simplex structure. Such a connection allows us to conveniently estimate the topic matrix from the output of SVD. 

Our contributions are several fold. First, the insight is new. In particular, our simplex is different from those in the literature (e.g., the simplicial cone by [8]), which are not based on the SVD.  
Second, we propose a new method for topic estimation, and it combines several ideas of complexity reduction. Its numerical performance is nice in our experiments. Last, our theoretical analysis is new and technically highly non-trivial, requiring delicate random matrix theory. 


From a high level, our work is related to \cite{SCC-JiJin,SCORE, mixed-SCORE}, which study social networks. The idea of taking entrywise ratios of eigenvectors was also used by \cite{SCORE} for network community detection. We study topic model estimation, so our work and those of \cite{SCC-JiJin, SCORE,mixed-SCORE} are motivated by different applications, and deal with different problems. So the results, including models, methods and theory, are all very different. 

We made an assumption that each topic has some anchor words. 
Such an assumption is largely for technical proofs and can be relaxed. For example, our results continue to hold as long as each topic has some ``nearly-anchor words". Numerically, we only require each topic to have very few ($<= 1\%$) anchor words; such an assumption is reasonable for many real examples. 



\section{Proof of Lemma~\ref{lem:key}}  \label{sec:proof}
In this lemma, we assume that all the entries of $\xi_1$ and $V_1$ are positive, so that the two matrices of entry-wise ratios, $R$ and $V^*$, are well-defined. 
Under mild conditions, the above assumption follows from the Perron's theorem; see the supplementary material.  

We first show that $r_i$ is a convex combination of $v_1^*,\cdots, v_K^*$ and that $\pi_i$ is the corresponding weight vector. By definition, $\Xi=AV=A[V_1,\cdots, V_K]$. It follows that $\xi_k(i)=a_i^TV_k$, for $1\leq k\leq K$. We plug this into the definition of $r_i$, $r_i(k)=\xi_{k+1}(i)/\xi_1(i)$, and obtain
\[
r_i(k)=\frac{a_i^TV_{k+1}}{a_i^T V_1} = \sum_{j=1}^K \frac{a_i(j)}{a_i^T V_1}\cdot V_{k+1}(j) = \sum_{j=1}^K \frac{a_i(j)V_1(j)}{a_i^T V_1}\cdot \frac{V_{k+1}(j)}{V_1(j)} = \sum_{j=1}^K \frac{a_i(j)V_1(j)}{a_i^T V_1} \cdot v_j^*(k). 
\]
By definition, $\Pi=[\mathrm{diag}(\xi_1)]^{-1} A\cdot\mathrm{diag}(V_1)$. So
\[
\pi_i(j) = \frac{a_i(j)V_1(j)}{\xi_1(i)} = \frac{a_i(j)V_1(j)}{a_i^TV_1}. 
\]
This implies that $\pi_i$ is indeed a weight vector. 
Combining the above, $r_i = \sum_{j=1}^K \pi_i(j)v_j^*$. So $r_i$ is a convex combination of $v_1^*, \cdots, v_K^*$ with weight $\pi_i$.

We then show that $v_1^*,\cdots,v_K^*$ form a non-degenerate simplex ${\cal S}^*_K$ by checking that $v_1^*,\cdots,v_K^*$ are affinely independent. 
By definition of $V^*$,  we have $[{\bf 1}_{K}, V^*]=V[\mathrm{diag}(V_1)]^{-1}$, where both $V$ and $\mathrm{diag}(V_1)$ are non-singular (since $V_1$ has all positive entries). So $[{\bf 1}_{K}, V^*]$ is non-singular, and it follows that the rows of $V^*$ are affinely independent. 

Last, it is seen that $r_i$ coincides with one vertex, i.e., $r_i=v_k^*$, if and only if $\pi_i(j)=0$ for all $j\neq k$, if and only if $a_i(j)=0$ for all $j\neq k$, if and only if word $i$ is an anchor word of topic $k$.

\bibliography{topic}
\bibliographystyle{imsart-number}

\end{document}